\documentclass[journal=jctcce,manuscript=article]{achemso}
\setkeys{acs}{articletitle = true}

\usepackage[version=3]{mhchem} 
\usepackage[T1]{fontenc}       
\usepackage{amsmath}
\usepackage[dvipsnames]{xcolor}

\author{Morgane Vacher}
\email{morgane.vacher@kemi.uu.se}
\author{Anders Brakestad}
\author{Hans O. Karlsson}
\author{Ignacio Fdez. Galván}
\author{Roland Lindh}
\email{roland.lindh@kemi.uu.se}
\affiliation[Uppsala University]
{Department of Chemistry -- \AA ngstr\"om, The Theoretical Chemistry Programme, Uppsala University, Box 518, 751 20 Uppsala, Sweden}

\title[]{Dynamical insights into the decomposition of 1,2-dioxetane}


\begin{document}

\begin{tocentry}
\includegraphics[scale=0.36]{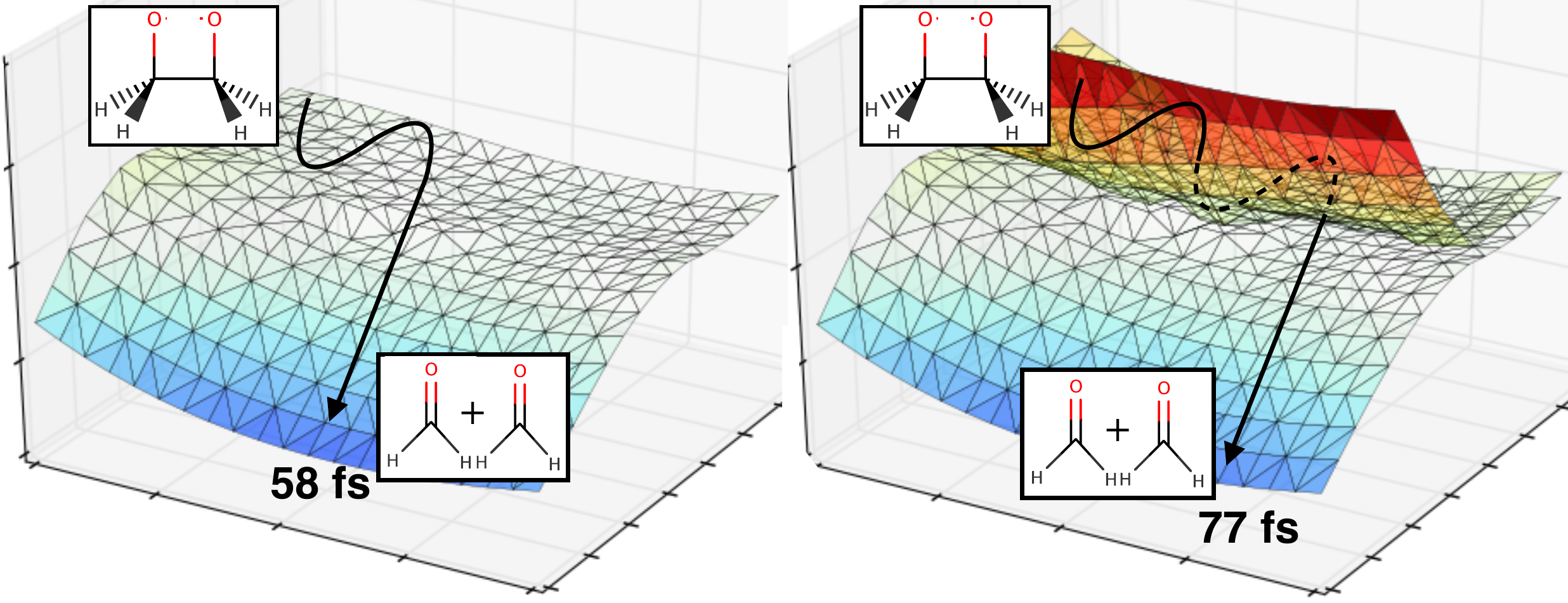}
\end{tocentry}

\begin{abstract}
Chemiluminescence in 1,2-dioxetane occurs via a thermally-activated decomposition reaction into two formaldehyde molecules. Both ground state and non-adiabatic dynamics (including singlet excited states) of the decomposition reaction have been simulated, starting from the first O--O bond breaking transition structure. The ground state dissociation occurs between t=30~fs and t=140 fs. The so-called entropic trap leads to frustrated dissociations, postponing the decomposition reaction. Specific geometrical conditions are necessary for the trajectories to escape from the entropic trap and for dissociation to be possible. The singlet excited states participate as well in the trapping of the molecule: dissociation including the non-adiabatic transitions to singlet excited states now occurs from t=30~fs to t=250~fs and later. Specific regions of the seam of $S_0$/$S_1$ conical intersections that would ``retain'' the molecule for longer on the excited state have been identified.
\end{abstract}

\section{Introduction}
Today's basic understanding of chemiluminescence is that a thermally activated molecule reacts and by doing so, it undergoes a non-adiabatic transition~\cite{Yarkony-2012,Yonehara-2012,Malhado-2014} up to an electronic excited state of the product, which then releases the excess energy in the form of light. In this respect, chemiluminescence can be seen as a reversed photochemical reaction: in photochemistry, a chemical reaction \textit{is caused} by the \textit{absorption} of light~\cite{Klessinger-1995,Bernardi-1996,Domcke-2012} while in chemiluminescence, a chemical reaction \textit{causes} the \textit{emission} of light~\cite{Turro-1973,Matsumoto-2004}. 
Nature has designed molecular systems with chemiluminescence properties~\cite{Navizet-2011}, the firefly luciferin--luciferase system being the most well-known example~\cite{Liu-2008,Navizet-2009,DaSilva-2011,Navizet-2013}. The process is often called bioluminescence when occurring in living organisms.

To understand and rationalise experimental observations, theoretical studies have investigated the detailed structure of the light emitting species\cite{Chen-2014,Cheng-2015,Tsarkova-2016}.
In the bacterial luciferase for instance, the light emitting species is a flavin derivative.
 Almost all currently known chemiluminescent systems have the peroxide bond --O--O-- in common. 
 The smallest system with chemiluminescent properties is the 1,2-dioxetane molecule.
 The general mechanism arising from previous works implies a stepwise process:~\cite{DeVico-2007,Farahani-2013} (i) the O--O bond is broken leading to a biradical region where four singlet states and four triplet states are degenerate and (ii) the C--C bond is broken leading to dissociation of the molecule into two formaldehyde molecules. For the last step, dark decomposition occurs if the two formaldehyde molecules are in the ground state while chemiluminescent decomposition occurs if one formaldehyde molecule ends up in a singlet or triplet excited state (Scheme~\ref{Scheme_Chemi}).
The efficiency of the chemiluminescent process is, however, observed to be low in 1,2-dioxetane. Besides, chemical titration of formaldehyde and chemiluminescence measurement show that the yield of the triplet excited states is much higher than that of the singlet excited states, and correspondingly, the phosphorescence yield is much higher than the fluorescence yield.~\cite{Adam-1985}

\begin{scheme}
  \includegraphics[scale=0.6]{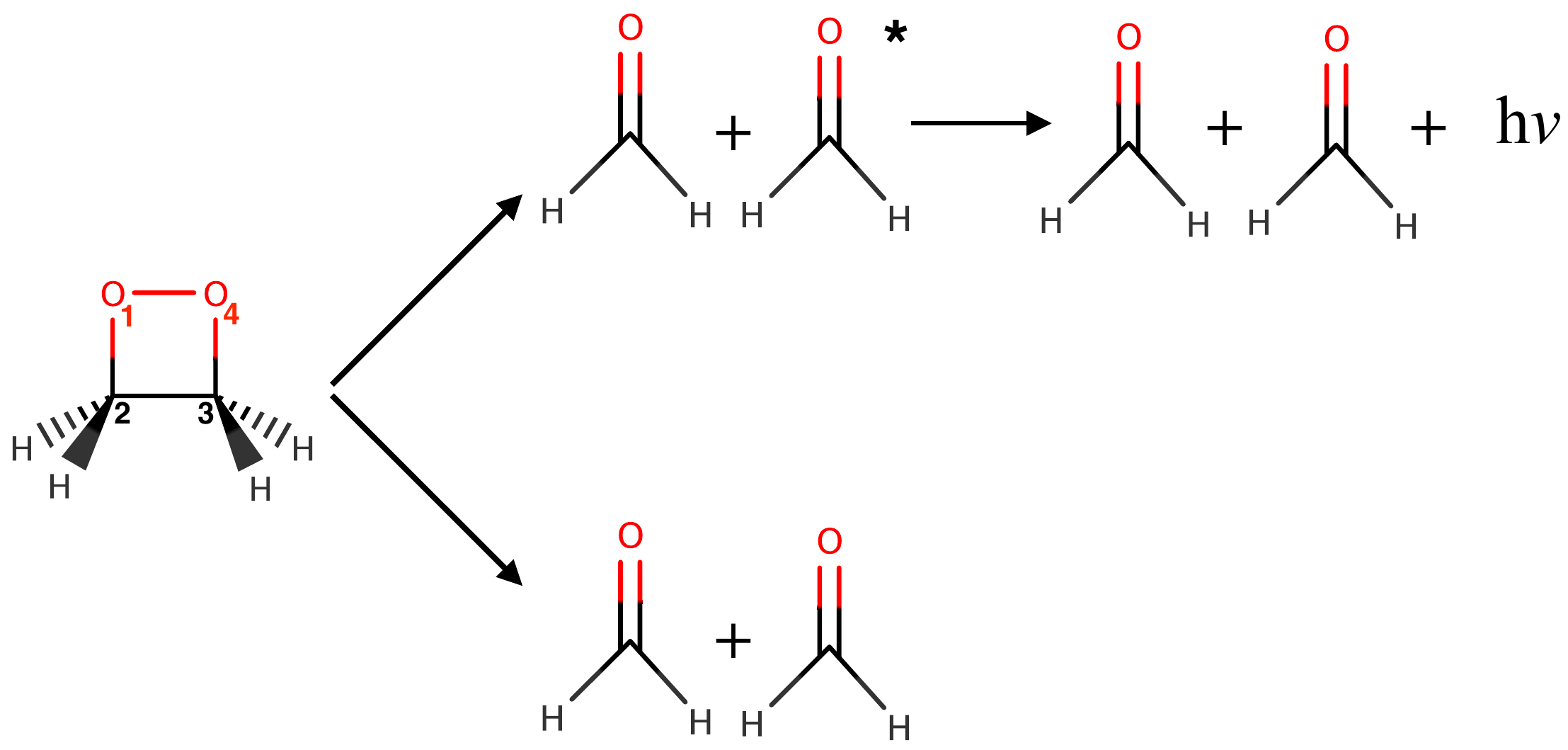}
  \caption{Dark and chemiluminescent decompositions of 1,2-dioxetane into two formaldehyde molecules (showing the numbering of C and O atoms in the reactant).}
  \label{Scheme_Chemi}
\end{scheme}

Previous theoretical studies on 1,2-dioxetane have investigated the reaction mechanism by computing cuts of potential energy surfaces and identifying critical points and potential pathways.~\cite{Reguero-1991,Wilsey-1999,Tanaka-2000,Sun-2012} The activation energy  calculated at the multi-state multi-configurational second order perturbation level of theory is in agreement with the experimentally observed value of 22 kcal/mol.~\cite{DeVico-2007,Farahani-2013} The favoured production of formaldehyde in the triplet excited states (compared to the singlet excited states) is explained by the difference in activation energies needed in order to break the C--C bond on the $S_1$ and $T_1$ potential energy surfaces. An interesting aspect of the computed paths is the suggested ``entropic trap''~\cite{DeFeyter-1999} that would regulate the outcome of the dissociation (instead of a transition state), by delaying the exothermic ground state dissociation and by instead giving the molecule the time to explore other routes and access the $S_1$ and $T_1$ surfaces for instance.

With simulations of the actual dynamics of the molecular system, i.e.\@ how the nuclei and electrons move as a function of time during the process, the present work provides new insights into this phenomenon and in particular about the ``entropic trap''. What defines it? How does it determine the efficiency of the chemiluminescence? What is the role of the singlet excited states? What is the lifetime in the biradical region? These are the questions addressed in the present article. By choosing to study the smallest system with chemiluminescent properties, the aim is to provide accurate and important predictions on the molecular basis of the reaction mechanism. The article is structured as follows: In section 2, the theoretical methods used and computational details are presented. In section 3, the results are presented and discussed. Section 4 offers some concluding remarks.

\section{Theoretical methods and computational details}
In this section, the theoretical methods used in the present work are presented, as well as the computational details. First, the initial conditions chosen for the dynamics simulations are described. Then, the on-the-fly semiclassical methods used for the dynamics and the electronic structure method used in combination with the former are presented briefly.

\subsection{Initial conditions}
The interest of the present work is in simulating the dynamics of the decomposition reaction of the 1,2-dioxetane molecule (in isolated conditions). An approach used to simulate post-transition state dynamics is to initialise the trajectories at a rate controlling transition state (TS), i.e.\@ here the TS for the O--O bond breaking, with Wigner sampling,\bibnote{Wigner and classical samplings were shown to give statistically same post-transition state dynamics, in contrast to photodissociation dynamics~\cite{Sun-2010}} and then propagate the dynamics from there.~\cite{Lourderaj-2008,Sun-2012,Farahani-2013} In this approach, it is assumed that: (i) recrossings of the TS are unimportant (which is supported by the dynamics simulations reported below), (ii) non-adiabatic electronic transitions are insignificant for the TS structure or earlier ones~\cite{Farahani-2013} and (iii) a Wigner distribution is maintained on the ground state before reaching the TS. By initiating the dynamics at the TS and giving a small amount of kinetic energy (1~kcal/mol) along the reaction coordinate towards the fragmentation products via the biradical region, it mimics a molecule that would come from the reactant conformation and the efficiency of the simulation is enhanced -- if the trajectories were started at the reactant conformation, substantial time would be used to propagate the motion from the reactant to the TS, and this would only happen in a small fraction of trajectories. Moreover, trajectories initiated in the vicinity of the reactant conformation would not have zero-point energy when they cross the TS.~\cite{Lourderaj-2008}
All the other normal modes are sampled (both positions and momenta) with a Wigner distribution to mimic the vibrational ground state along these other coordinates, using the Newton-X package~\cite{Newton-X}. It can be noted that for the normal mode with the lowest (real) frequency, the ratio between the vibrational first excited and ground state populations is less than $1/3$ according to the Boltzmann distribution at $T=300$ K; for the normal mode with the second lowest (real) frequency, it is less than $1/10$. The population of vibrational excited states can therefore be safely neglected and only the vibrational ground state is sampled.
Note that these initial conditions have been chosen as a way to explore the mechanism of the dissociative reaction. For direct comparison with experiments performed at a certain temperature~$T$, the kinetic energy along the reaction coordinate should also be sampled so that its average value is $RT$. This is not done in the present work.

\subsection{On-the-fly semiclassical dynamics}
``On-the-fly'' dynamics is performed, i.e.\@ the potential energy surfaces are calculated as needed along nuclear trajectories. The nuclear motion is treated classically integrating numerically Newton's equations of motion using the velocity Verlet algorithm. A time step of 10~a.u., i.e.\@ about 0.24~fs is used. All nuclear coordinates are taken into account. Both Born--Oppenheimer adiabatic dynamics and non-adiabatic dynamics using the surface hopping method with the Tully's fewest switches algorithm~\cite{Tully-1971,Tully-1990,Malhado-2014} are performed. In the former simulations, the molecule is bound to stay in the electronic ground state while in the latter, transitions among any on the four lowest-energy singlet states are included. The decoherence correction proposed by Granucci and Persico is used with a decay factor of 0.1~hartree.~\cite{Granucci-2007} The implementation of the above methods in a development version of the Molcas package is used.~\cite{Molcas-2016}

\subsection{Multi-reference electronic structure}
Regarding the electronic structure, the complete active space self-consistent field (CASSCF)~\cite{Roos-1980,Roos-1987} method state-averaging over the four singlet states equally is used to optimise the TS, calculate the normal modes at the TS structure, and calculate the energy and its first derivative with respect to the nuclear distortions along the nuclear trajectories. The active space used consists of 12~electrons distributed in 10~orbitals: the four $\sigma$ and four $\sigma^*$ orbitals of the four-membered ring, plus the two oxygen lone-pair orbitals perpendicular to the ring. The ANO-RCC basis set with polarised triple-zeta contraction (ANO-RCC-VTZP)~\cite{Roos-2004}\bibnote{ANO basis sets are systematically improvable, and produce smaller basis set superposition errors. Also, the ANO-RCC basis set was developed using correlated methods and aimed at providing a balanced description of various electronic states.} and the atomic compact Cholesky decomposition (acCD)~\cite{Aquilante-2009} are used.

It can be noted that previous theoretical works have shown that a multi-configurational method with second order perturbation correction (together with ANO-RCC-VTZP basis set) was needed for quantitative agreement of the activation energy of the O--O bond breaking with the experimental value.~\cite{DeVico-2007} However, the second order perturbation correction does not seem necessary for the quantitatively good relative potential energies beyond the first TS (see Supporting Information).~\cite{DeVico-2007} It is therefore considered that the dynamics after the O--O TS can be safely studied at CASSCF level of theory.

\section{Results and discussion}
In this section, the results of the simulations are presented. First, the adiabatic ground state dynamics simulations are discussed and then the non-adiabatic dynamics simulations including the singlet excited states. By separating the results this way, the role of the singlet excited states is isolated.

\subsection{Ground state adiabatic dynamics}
First, the result of a single ground state dynamics simulation (with no initial Wigner sampling of nuclear positions and velocities) is presented in order to qualitatively describe the nuclear motion during the decomposition reaction. Then, the results of an ensemble of trajectories are presented for quantitative information.

\subsubsection{Result of a single un-sampled trajectory}
A single trajectory has been started at the O--O bond breaking TS with 1~kcal/mol of kinetic energy along the transition vector towards the biradical product: this is the ``un-sampled'' trajectory. The transition vector at the TS consists of mainly a rotation around the C--C bond, i.e.\@ an increase of the O$_1$-C$_2$-C$_3$-O$_4$ dihedral angle, together with a symmetric increase of the O$_1$-C$_2$-C$_3$ and O$_4$-C$_3$-C$_2$ angles, therefore breaking the O--O bond (see Scheme~\ref{Scheme_Chemi} for the atom numbering).

\begin{figure}
  \includegraphics[scale=0.58]{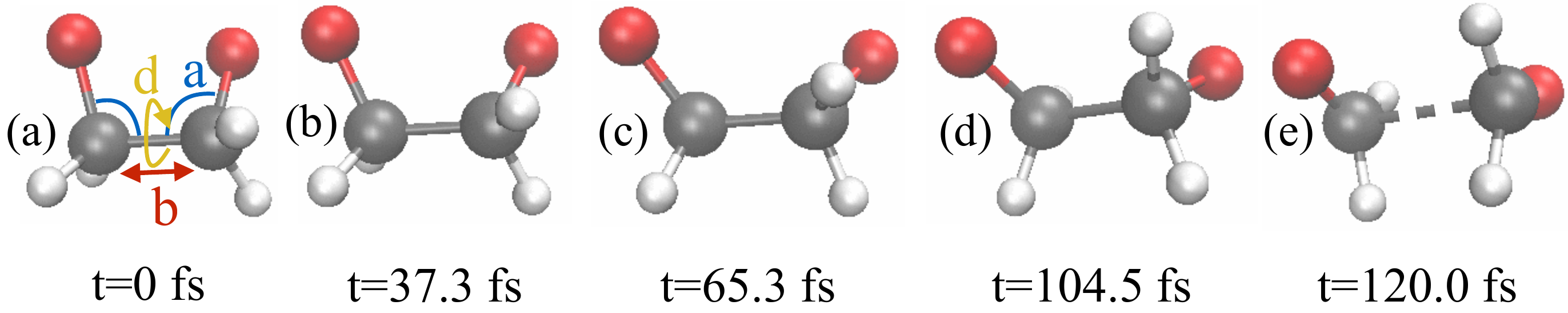}
  \caption{Snapshots of the nuclear geometry during the un-sampled trajectory. The relevant C--C bond, the O-C-C angles and the O-C-C-O dihedral angle are indicated with `b', `a' and `d', respectively, in the snapshot (a). The geometrical parameters at the TS (a) are: b=1.5~\AA, a=101$^\circ$ and d=37$^\circ$. The C--C bond length reaches 2.4~\AA~at approximately 120~fs. (Multimedia view available online)}
  \label{Fig_Mol}
\end{figure}

\begin{figure}
  \includegraphics[scale=1]{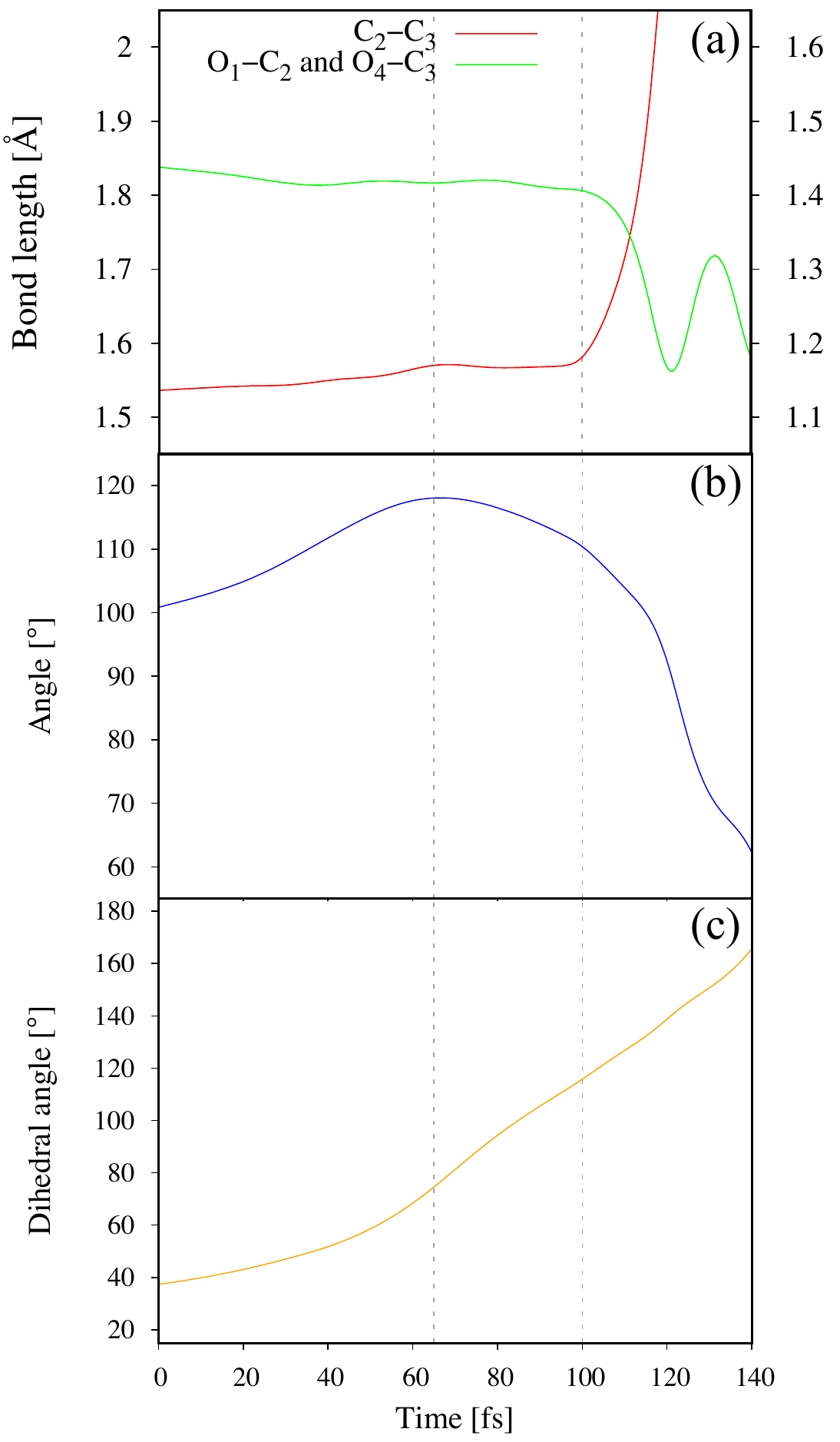}
  \caption{Evolution of the (a) C$_2$-C$_3$ bond (left $y$ axis), O$_1$-C$_2$ and O$_4$-C$_3$ bonds (right $y$ axis), (b) O$_1$-C$_2$-C$_3$ and O$_4$-C$_3$-C$_2$ angles and (c) O$_1$-C$_2$-C$_3$-O$_4$ dihedral angle during the un-sampled trajectory. The two O-C bonds and the two O-C-C angles stay equal during the trajectory because of the symmetry of the molecule. The dashed lines indicate the time window during which the C$_2$-C$_3$ bond stops increasing and stays roughly constant. See Scheme~\ref{Scheme_Chemi} for the atom numbering.}
  \label{Fig_Dyn}
\end{figure}

Figure~\ref{Fig_Mol} shows snapshots of the nuclear geometry during the un-sampled trajectory. The evolution of the bond lengths, angles and dihedral angle are plotted in Figure~\ref{Fig_Dyn}.
The C--C bond lengthens slightly up to t=65~fs, then stays roughly constant until t=100~fs when it suddenly dissociates (red curve in Figure~\ref{Fig_Dyn}a). It reaches 2.4~\AA~(two times the Van der Waals radius of a carbon atom) at approximately 120~fs. The two C--O bonds stay equal during the trajectory because of the symmetry of the molecule. They suddenly decrease after t=100~fs from 1.4~\AA~(typical value for a single bond) to 1.2~\AA~(typical value for a double bond); the bonds actually oscillates around 1.2~\AA~because of the excess of kinetic energy (green curve in Figure~\ref{Fig_Dyn}a). In the meantime, the dihedral angle increases monotonically from 37$^\circ$ (value at the TS) to about 140$^\circ$ at t=120~fs (Figure~\ref{Fig_Dyn}c). The initial increase in the O$_1$-C$_2$-C$_3$-O$_4$ dihedral angle is expected since the dihedral angle is part of the transition vector. Similarly, the two O-C-C angles (equal because of the symmetry of the molecule) initially increase, but they reach a maximum of about 118$^\circ$ at t=65~fs and then decrease back (Figure~\ref{Fig_Dyn}b). Why do the O-C-C angles decrease at some particular time? Why does the C--C bond stop increasing at this same time as if something was holding back the dissociation?

\begin{figure}
  \includegraphics[scale=0.93]{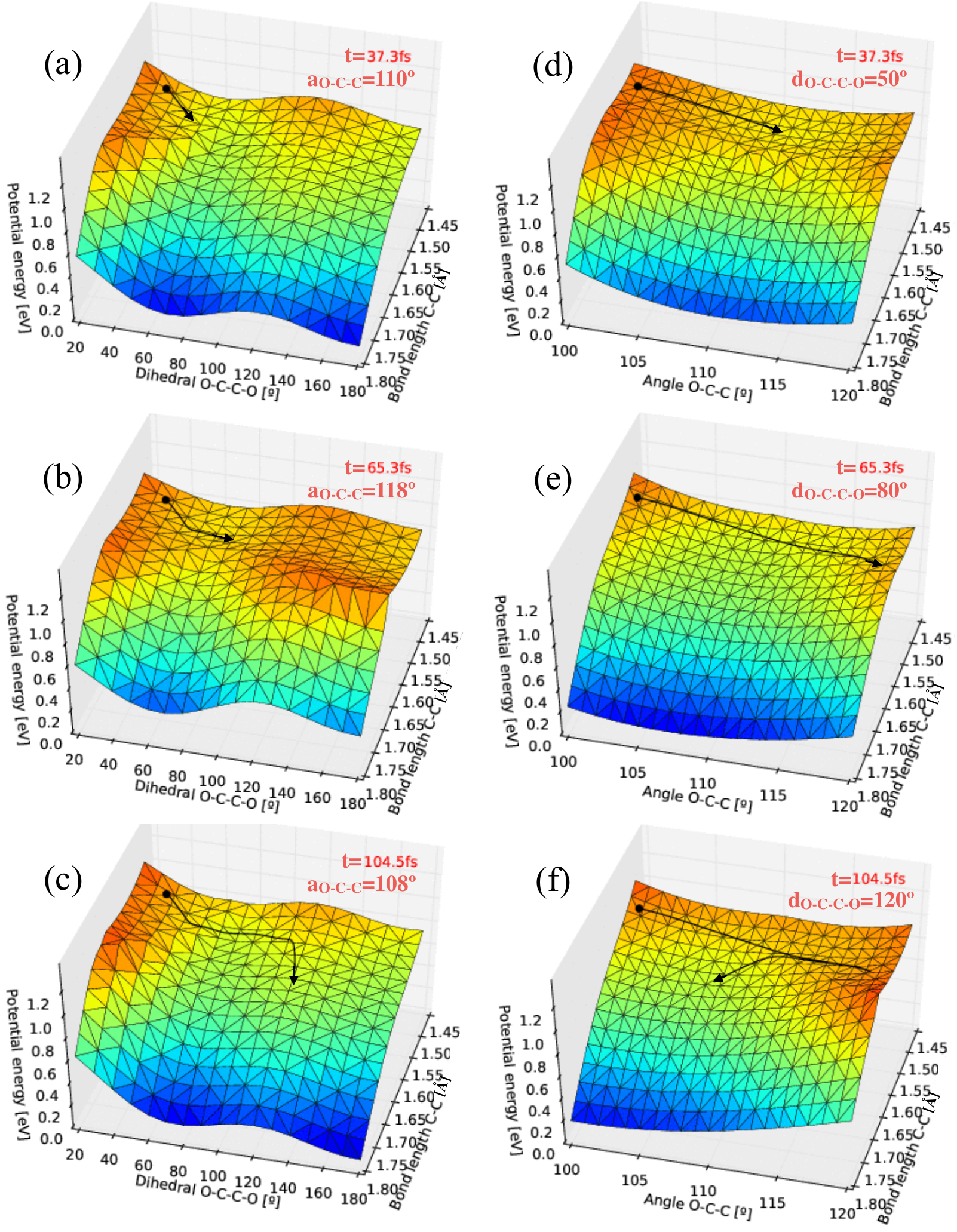}
  \caption{Potential energy surfaces of $S_0$ felt during the un-sampled trajectory at (a,d) t=37.3~fs, (b,e) 65.3~fs and (c,f) 104.5~fs. The black arrow shows the evolution of the un-sampled trajectory. \textbf{Left side panels:} The nuclear coordinates are the O-C-C-O dihedral angle and C--C bond length for O-C-C angles of (a) 110$^\circ$, (b) 118$^\circ$ and (c) 108$^\circ$. \textbf{Right side panels:} The nuclear coordinates are the O-C-C angles and C--C bond length for O-C-C-O dihedral angles of (d) 50$^\circ$, (e) 80$^\circ$ and (f) 120$^\circ$. The other nuclear coordinates were optimised to minimise the potential energy.  (Multimedia view available online)}
  \label{Fig_Scan}
\end{figure}

In order to understand and rationalise these observations, the ground state potential energy surfaces ``felt'' along the un-sampled trajectory at t=37.3~fs, 65.3~fs and 104.5~fs, i.e. times prior dissociation, have been studied. Figure~\ref{Fig_Scan} shows, for each of these three times, 2-dimensional cuts of the $S_0$ potential energy surfaces with either the O-C-C angles (left side panels) or the O-C-C-O dihedral angle (right side panels) fixed to their respective values at these times during the un-sampled trajectory, while scanning either the C--C bond length / O-C-C-O dihedral angle (left side panels) or C--C bond length / O-C-C angles (right side panels) -- see Figure~\ref{Fig_Mol}a for the selected nuclear coordinates. More precisely, the C--C bond length is scanned from 1.5~\AA~to 1.8~\AA~with a step of 0.02~\AA, the O-C-C angles (maintaining them equal) from 100$^\circ$ to 120$^\circ$ with a step of 1$^\circ$, and the O-C-C-O dihedral angle from 20$^\circ$ to 180$^\circ$ with a step of 10$^\circ$. The other nuclear coordinates were optimised to minimise the potential energy. 
Initially, the O-C-C-O dihedral angle, the O-C-C angles and the C--C bond length increase (Figures~\ref{Fig_Scan}a and \ref{Fig_Scan}d). At around t=65~fs, a barrier towards dissociation appears, forcing the nuclear trajectory to stay at short C--C bond lengths, while the dihedral angle keeps on increasing (Figure~\ref{Fig_Scan}b); the O-C-C angles reach their maximum value since higher angles are unfavourable (Figure~\ref{Fig_Scan}e). Finally, the barrier disappears when the O-C-C angles decrease back to a lower value and the molecule dissociates (Figures~\ref{Fig_Scan}c and \ref{Fig_Scan}f). In summary, the single un-sampled trajectory and the 3-dimensional scan show a barrier towards dissociation at low dihedral angles (<60$^\circ$), and also at large O-C-C angles (>115$^\circ$).

\subsubsection{Result of an ensemble of trajectories}
In this subsection, the results of an ensemble of 150 ground state trajectories are presented in order to give quantitative information about the dynamics of the decomposition reaction. (150 trajectories are enough to obtain a converged result in the present case, as demonstrated in Supplementary Information.) It is noted that barrier recrossings occur sometimes during chemical reactions. Here, none of the 150 trajectories initially directed towards the product recrossed the TS. TS recrossing is therefore assumed to be unimportant for this specific reaction.

\begin{figure}
  \includegraphics[scale=0.6]{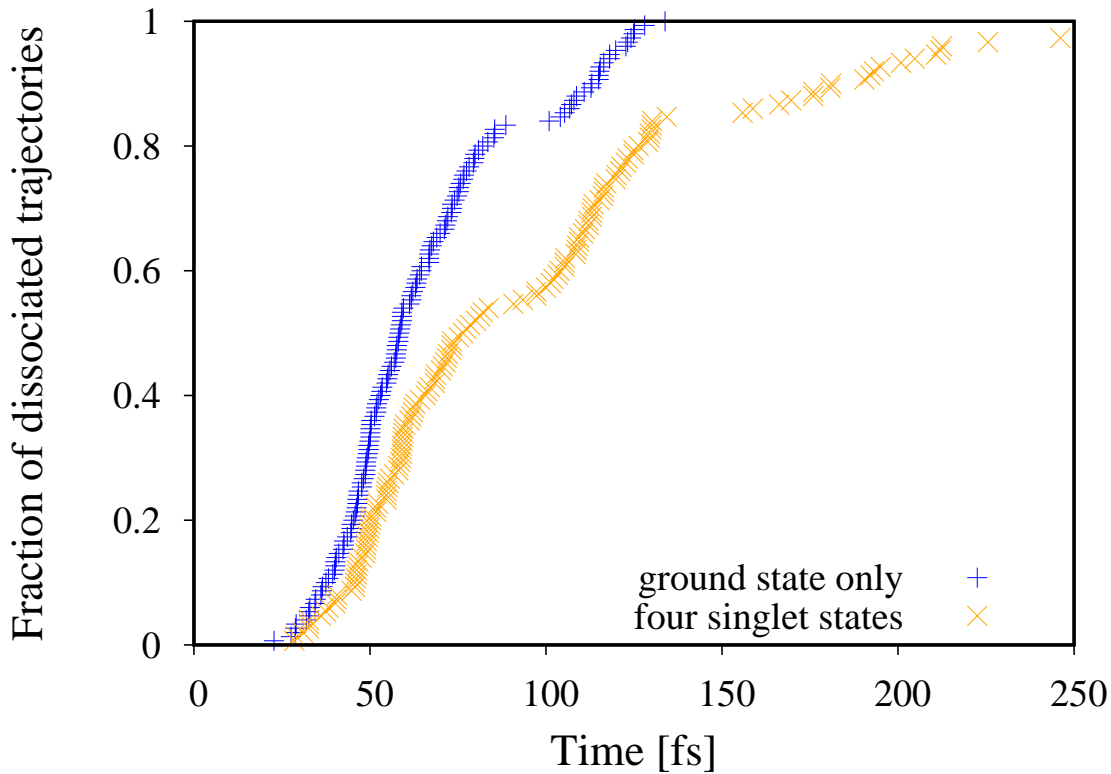}
  \caption{Dissociation time scale of an ensemble of 150 ground state trajectories (blue) and surface hopping trajectories including the four lowest-energy singlet states (orange), both started from a Wigner sampling. Dissociation is considered to occur when the C--C bond length exceeds 2.4~\AA.}
  \label{Fig_DynTime}
\end{figure}

Figure~\ref{Fig_DynTime}~shows the time evolution of the fraction of ground states trajectories that have dissociated (blue curve). Dissociation is considered to occur when the C--C bond length exceeds 2.4~\AA~(two times the Van der Waals radius of a carbon atom). It requires approximately 30~fs for the first trajectories to dissociate, and approximately another additional 30~fs for half of the trajectories to dissociate: t$_{1/2}^{BO}$=58~fs. In less than 150~fs, all the trajectories have dissociated. The existence of a gap in the dissociation times can be noted around t=95~fs. Here, the ``arbitrariness'' of the value of 2.4~\AA~should be commented briefly. Choosing a slightly smaller (or larger) value leads to a shift of the dissociation time curve towards slightly earlier (or later) times (see Supplementary Information); but it does not change any relative comparison, nor the conclusions.

\begin{figure}
  \includegraphics[scale=0.8]{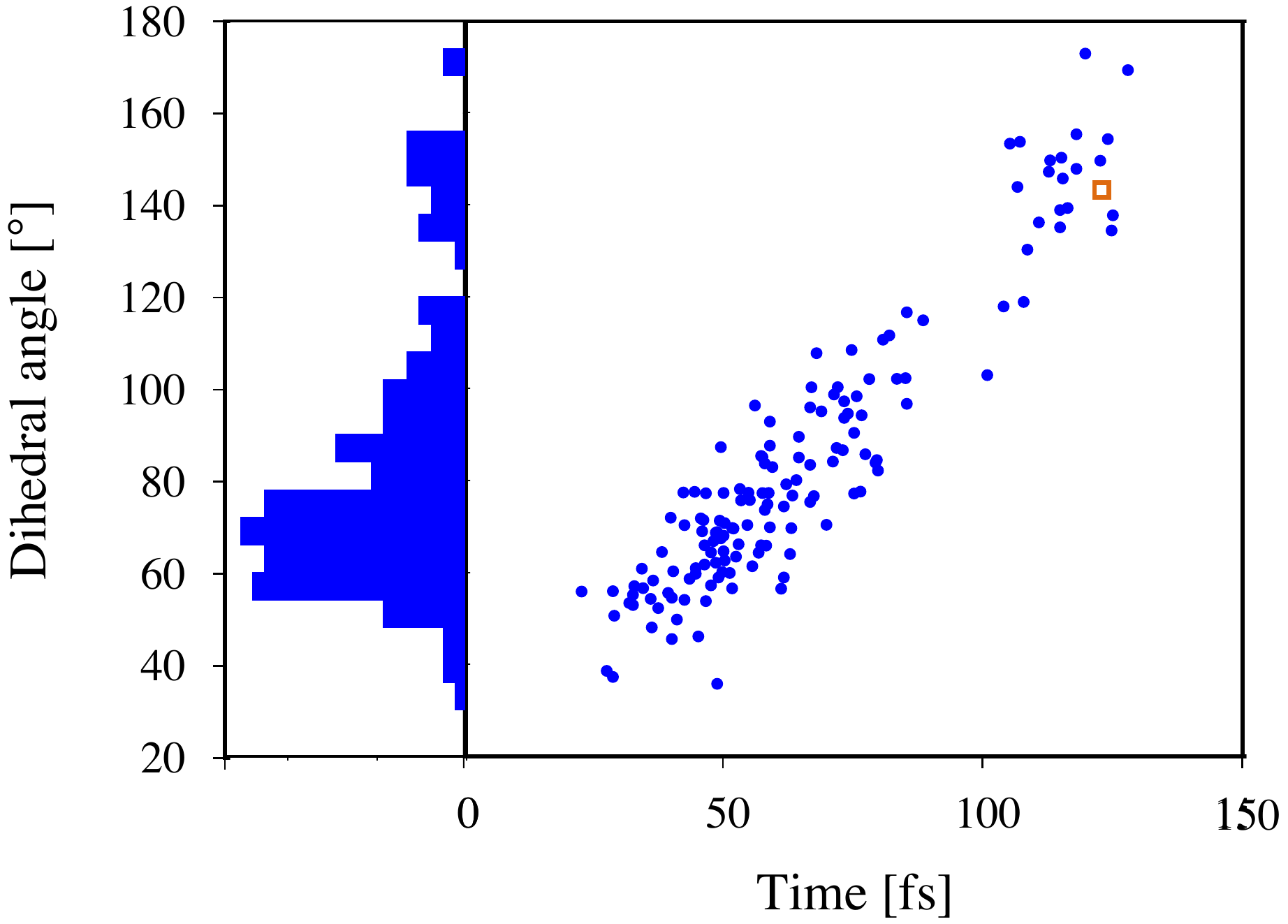}
  \caption{Correlation between dissociation time and the O-C-C-O dihedral angle at dissociation for the ensemble of 150 ground state trajectories. The orange square represents the result of the single un-sampled trajectory. The histogram on the left corresponds to the distribution of dihedral angles at dissociation. Dissociation is considered to occur when the C--C bond length exceeds 2.4~\AA.}
  \label{Fig_DihTime}
\end{figure}

In Figure~\ref{Fig_DihTime}, the O-C-C-O dihedral angles at dissociation are plotted against the times of dissociation for the ensemble of 150 ground state trajectories. There is a clear correlation between the two quantities: as the trajectory takes longer to dissociate, the dihedral angle has time to reach larger values. Also, the gap in dissociation times corresponds to a gap in dihedral angles at dissociation around 120$^\circ$. This can be understood with the shape of the potential energy surface: the dihedral angle of 120$^\circ$ is a maximum along the dihedral angle coordinate (Figures~\ref{Fig_Scan}a-c). If a trajectory has enough kinetic energy along the dihedral angle coordinate to go over the maximum, one expects it to carry on a bit further towards larger dihedral angles and not make a 90$^\circ$ turn at the maximum and dissociate on the ridge. It is therefore more likely to dissociate at a dihedral angle smaller or larger than 120$^\circ$, hence the gap in dihedral angles at dissociation and the corresponding gap in dissociation times. In Figure~\ref{Fig_DihTime}, the result of the single un-sampled trajectory is indicated with an orange square: dissociation occurs at t=120~fs with a dihedral of 140$^{\circ}$.  Although the single trajectory is consistent with the results of the ensemble of trajectories, it is here not representative of the ensemble. Most trajectories dissociate earlier than the un-sampled one, due to the additional kinetic energy (from the zero-point energy) especially along the C--C bond coordinate. This demonstrates the importance of running an ensemble of trajectories with sampled initial conditions.

\begin{figure}
  \includegraphics[scale=0.6]{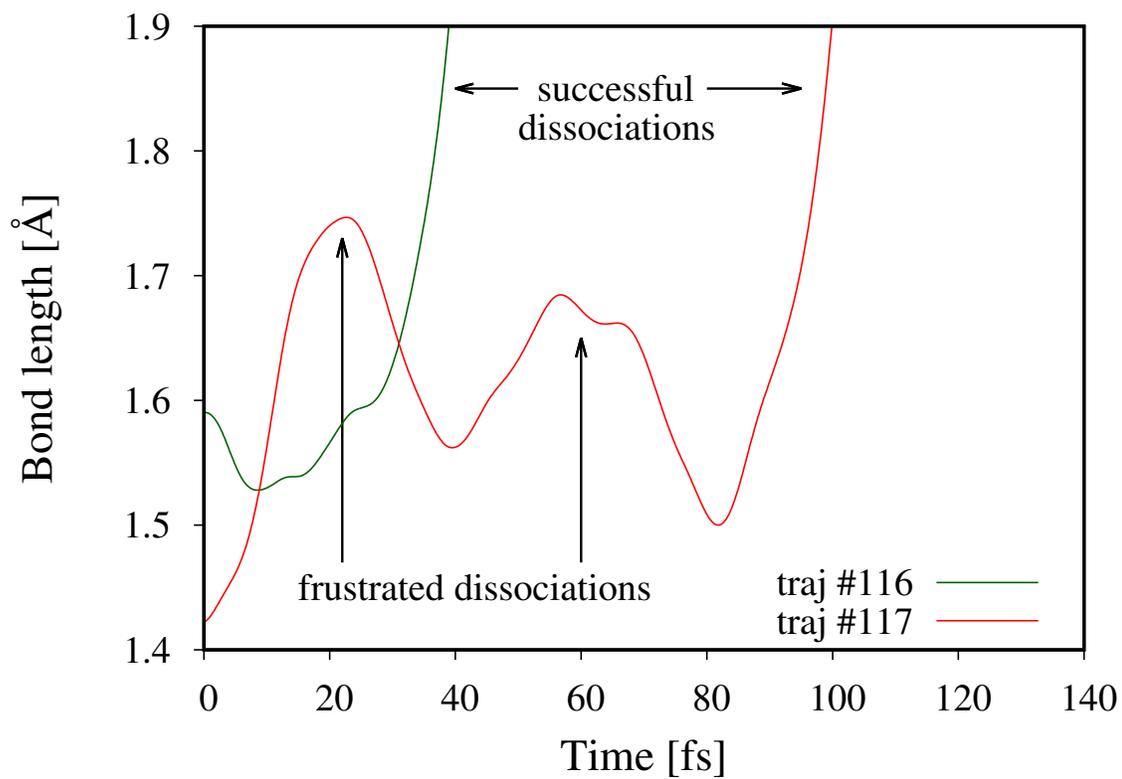}
  \caption{Evolution of the C$_2$-C$_3$ bond length during two representative trajectories of the ensemble.}
  \label{Fig_BondSampl}
\end{figure}

\begin{figure}
  \includegraphics[scale=0.6]{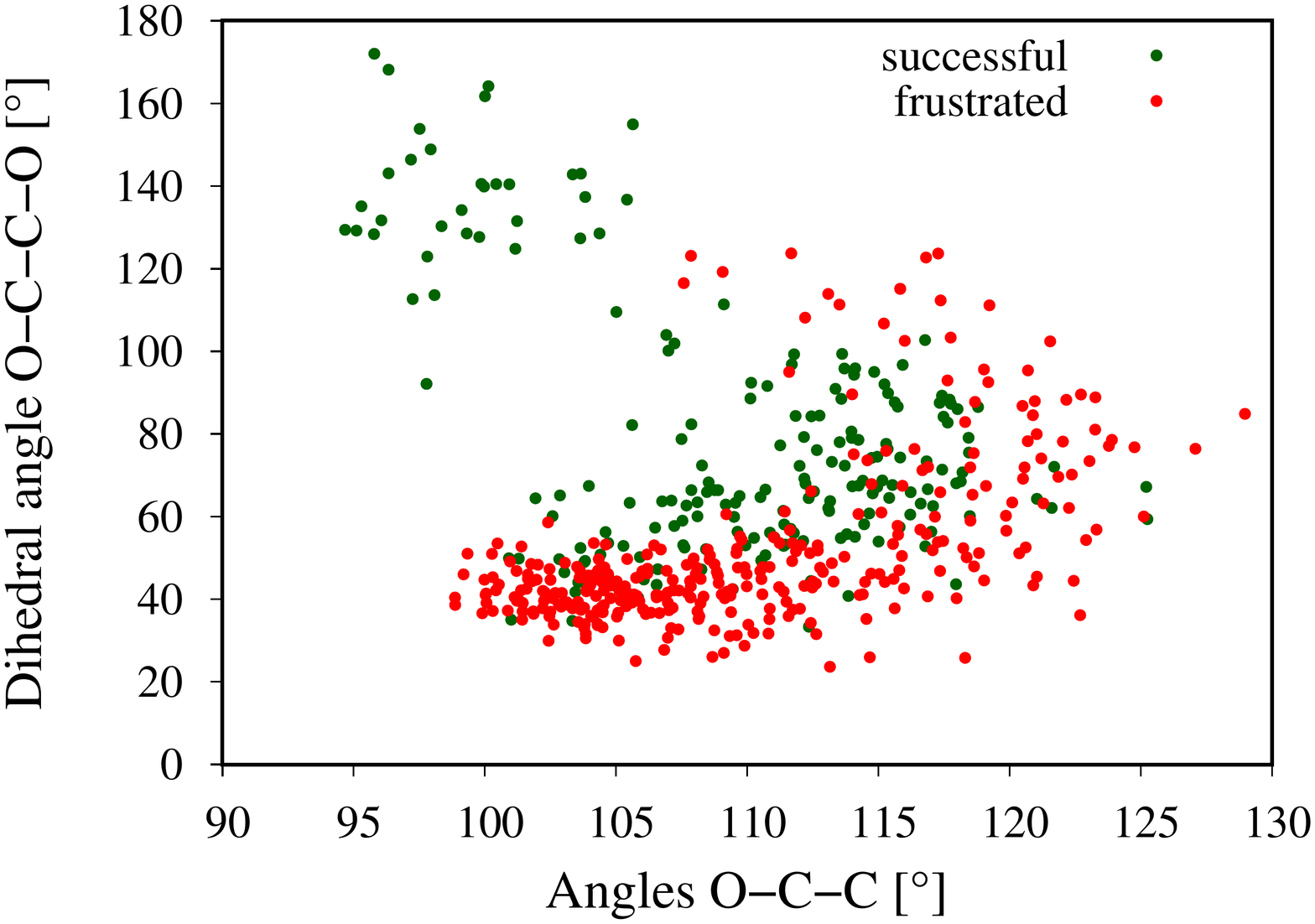}
  \caption{Geometrical characteristics of the frustrated (red) and successful (green) dissociations of the C--C bond studied over the ensemble of 150 ground state trajectories. Frustrated dissociations are identified by maxima of the C--C bond length time evolution.}
  \label{Fig_Frus}
\end{figure}

The distribution of dihedral angles at dissociation is explained by the distribution of dissociation times due to the simple linear correlation. But what explains the distribution of dissociation times? Why do some trajectories dissociate earlier than others? Figure~\ref{Fig_BondSampl} shows the evolution of the C$_2$-C$_3$ bond length during two representative trajectories of the ensemble: trajectory \#116 that dissociates at t=45~fs and trajectory \#117 that exhibits two ``frustrated'' dissociations before finally dissociating at t=105~fs. A longer dissociation time scale is thus due to these frustrated dissociations. It can be noted in passing that, in trajectory \#117, the C--C bond oscillates with a period of approximately 35~fs, which is typical of C--C vibrations. In order to rationalise the existence of frustrated dissociations, the characteristics of the geometries at which dissociations are frustrated have been studied in comparison to those at which dissociations are successful over the ensemble of trajectories (Figure~\ref{Fig_Frus}).~\bibnote{Frustrated dissociations lead to elongation of the C--C bond up to 1.8~\AA. To make the comparison between the geometries of frustrated versus successful dissociations fair, dissociation is here considered to happen when the C--C bond length exceeds 2.0~\AA~(instead of 2.4~\AA).} Frustrated dissociations are identified by the fact that the C--C bond length time evolution presents a maximum when this happens. The following trends are observed: (i) most dissociations that are attempted at dihedral angles lower than 55$^\circ$ are frustrated and (ii) the same for dissociations that are attempted at O-C-C angles larger than 117$^\circ$. These results are again consistent with the shape of the potential energy surface of the electronic ground state, the characteristics of which are highlighted by the 3-dimensional scan presented in the previous subsection (Figure~\ref{Fig_Scan}).

To end this section, the present results are compared with a previous theoretical study where ground state dynamics in 1,2-dioxetane was performed.~\cite{Farahani-2013} There, the simulations were run with a thermostat at a constant temperature of 300~K. The initial conditions of the trajectories were also different: no distortion of the geometry was done and the velocities were sampled to reproduce a Maxwell--Boltzmann distribution at 300~K. A dissociation half-life of 613~fs was then obtained. The much shorter dissociation time scale found in the present work (58~fs) can be explained by the fact that the trajectories are started at the O--O bond breaking transition structure and ``pushed'' towards the biradical product (which was not done in the previous study). It is noted that ground state direct dynamics simulations in 1,2-dioxetane were also performed previously in an another work,~\cite{Sun-2012} investigating the formation from O$_2$ and ethylene, and the decomposition. In that study at UB3LYP/6-31G$^*$ level of theory, the trajectories were initiated at a different TS; few of those that formed the dioxetane molecule then dissociated into two formaldehyde molecules.

\subsection{Non-adiabatic dynamics with singlet excited states}
\begin{figure}
  \includegraphics[scale=0.6]{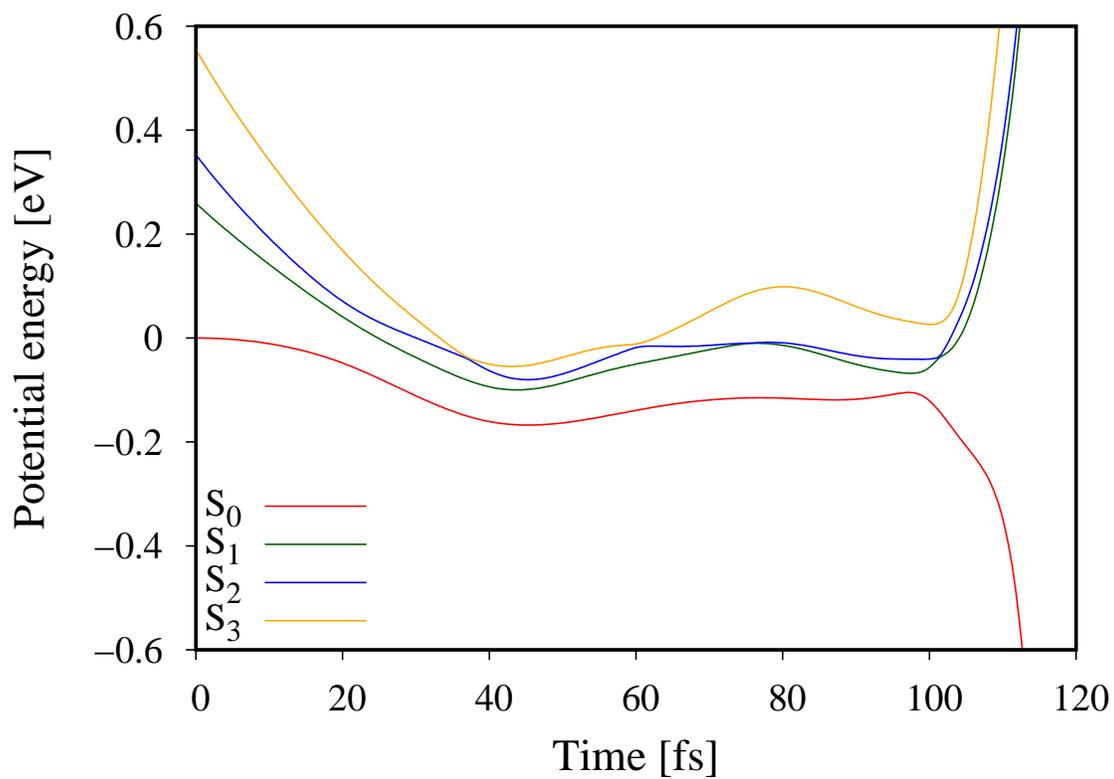}
  \caption{Potential energies of the four lowest-energy singlet states along the un-sampled ground state trajectory.}
  \label{Fig_Ene}
\end{figure}

In this section, the role of the three singlet excited states on the decomposition reaction is now investigated. Before presenting the results of surface hopping trajectories, Figure~\ref{Fig_Ene} shows the potential energies of the four lowest-energy singlet states along the un-sampled trajectory on the electronic ground state. The potential energy of $S_0$ drops after t=100~fs when the molecule dissociates (red curve) but during the reaction, the four singlet states come very close in energy (within 0.1~eV) and stay very close in energy for an extended period of time up to dissociation.\bibnote{In a previous theoretical study~\cite{Luk-2015}, where the photoisomerisation of different rhodopsins was investigated by simulating the dynamics on the first excited state, it was suggested that near degeneracy with the second electronic excited state delays the non-radiative transition back to the ground state. Similarly, in the present study, the near degeneracy with the singlet excited states could be responsible for the time it takes for the molecule to dissociate. Here, the population of these excited states is even allowed using non-adiabatic dynamics.}
Crossing points among the singlet excited states can even be seen. Internal conversions, i.e. non-adiabatic transitions among the singlet states, are therefore expected. Dissociation on a singlet excited state would eventually lead to luminescence. It is noted that four triplet states are also lying close in energy (not shown) and intersystem crossings between the singlet and triplet states could occur because of spin-orbit coupling. In order to study the effect of the singlet excited states, the results of an ensemble of 150 surface hopping trajectories -- where non-adiabatic transitions among the four singlet states are allowed -- are presented in this section. The initial conditions are the same as for the ground state dynamics simulations. The analysis is focussed on the differences with the ground state dynamics results.

Figure~\ref{Fig_DynTime}~shows the time evolution of the fraction of surface hopping trajectories that have dissociated (orange curve). It has a similar shape as the time evolution of the ground state trajectories (blue curve) with the same gap in the dissociation times around t=90~fs that correlates again with a gap in the dihedral angles at dissociation around 120$^\circ$ (not shown); but there is in addition a ``tail'' of longer-lived trajectories that dissociate after t=150~fs. This increases the dissociation half-time to t$_{1/2}^{S}$=77~fs, compared to t$_{1/2}^{BO}$=58~fs when neglecting the non-adiabatic transitions. 

\begin{figure}
  \includegraphics[scale=0.6]{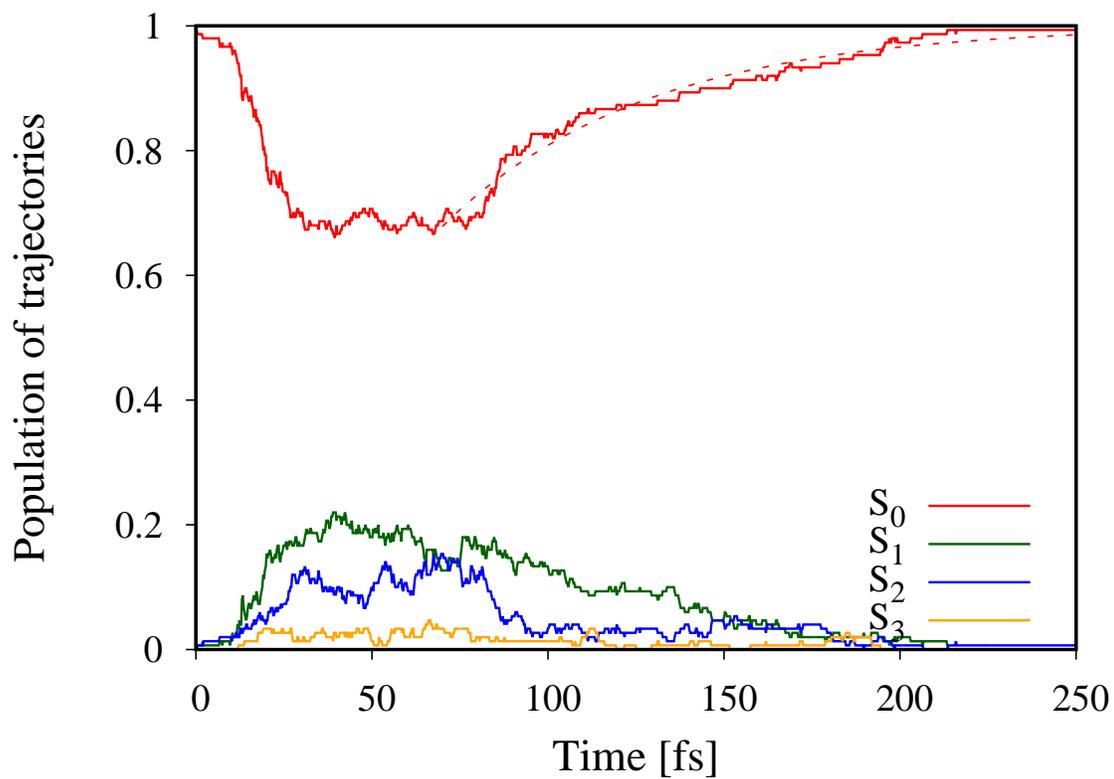}
  \caption{Electronic state populations of the ensemble of 150 surface hopping trajectories (including the four lowest-energy singlet states). The dashed red line corresponds to the function $P(t)=1-\exp (- (t-t_{\text{off}})/{\tau} )$ fitted to the population of the $S_0$ state after t=70~fs: a rise time of $\tau$=58~fs is obtained.}
  \label{Fig_Pop}
\end{figure}

Figure~\ref{Fig_Pop} shows the time evolution of the electronic state populations of the ensemble of trajectories. The total number of hops (or non-adiabatic transitions) among any of the four singlet states is on average more than 4.7 per trajectory. This illustrates the significant non-adiabatic character of the decomposition process. 
After 25~fs only, the population on $S_0$ drops to 0.70. Both $S_1$ and $S_2$ states are significantly populated, $S_1$ being about twice as populated as $S_2$, while $S_3$ gets almost no population. The ground state starts to be re-populated after t=70~fs on a time scale of approximately $\tau$=58~fs (dashed red line).~\bibnote{The fact that the rise time $\tau$ is equal to the ground state dissociation half-life t$_{1/2}^{BO}$ is obviously a coincidence.}
Importantly, no dissociation on the singlet excited state was obtained, which is expected because of the extremely low excitation yield in 1,2-dioxetane decomposition (0.0003 \% of formaldehyde singlet excited states).~\cite{Adam-1985} The present result is thus consistent with the experimental observation.

\begin{figure}
  \includegraphics[scale=0.6]{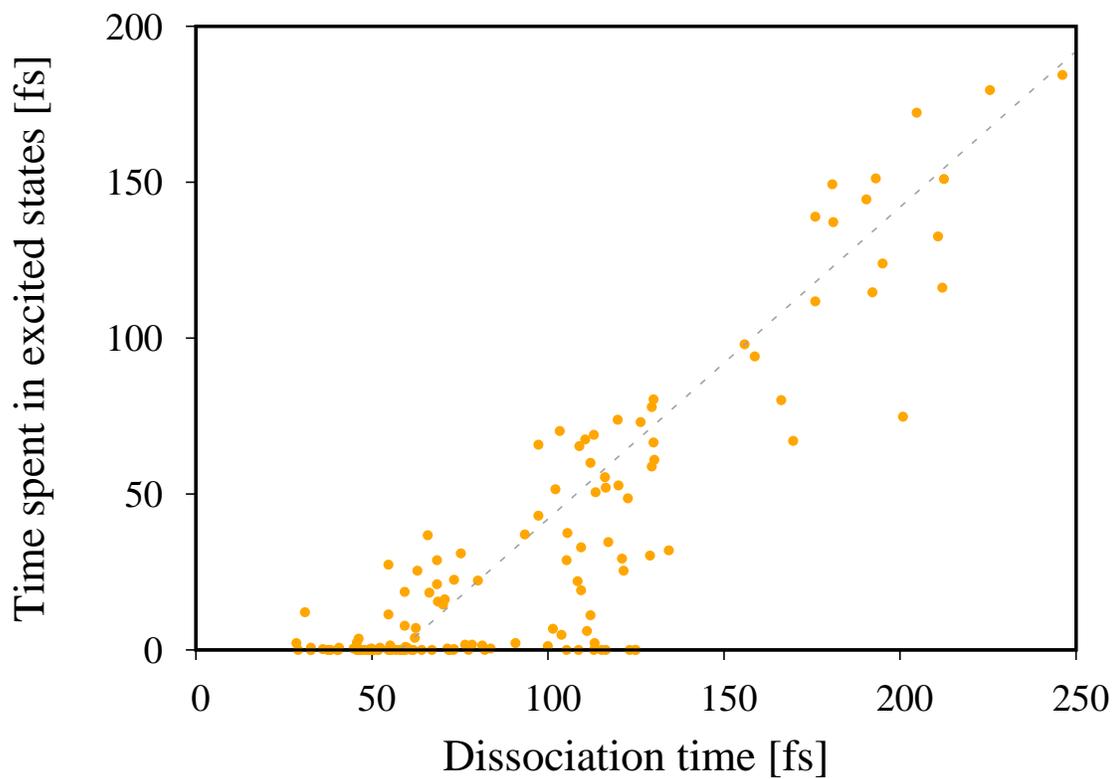}
  \caption{Correlation between dissociation time and the time spent in singlet excited states for the ensemble of 150 surface hopping trajectories. Dissociation is considered to occur when the C--C bond length exceeds 2.4~\AA. The dashed line has a slope of 1 and intersects the $x$ axis at t$_{1/2}^{BO}$=58~fs.}
  \label{Fig_Exc}
\end{figure}

In Figure~\ref{Fig_Exc}, the times spent in the singlet excited states are plotted against the times of dissociation for the ensemble of 150 surface hopping trajectories. The clear correlation between the two quantities is illustrated by the dashed line: 
the latter has a slope of 1 and intersects the $x$ axis at t$_{1/2}^{BO}$=58~fs (the dissociation half-time for Born--Oppenheimer ground state dynamics). The time spent in the singlet excited states seems to be effectively simply added to the ground state dissociation time in order to give the total time for dissociation (with non-adiabatic transitions).
It looks as if the dissociation clock was paused when the trajectory hopped up to the excited state and the clock resumed when the trajectory hopped back down.
As the trajectory stays longer in the singlet excited states, it takes longer for the molecule to dissociate. 
So the singlet excited states participate in the ``trapping'' of the molecule, by postponing the dissociation further.

\begin{figure}
  \includegraphics[scale=1.0]{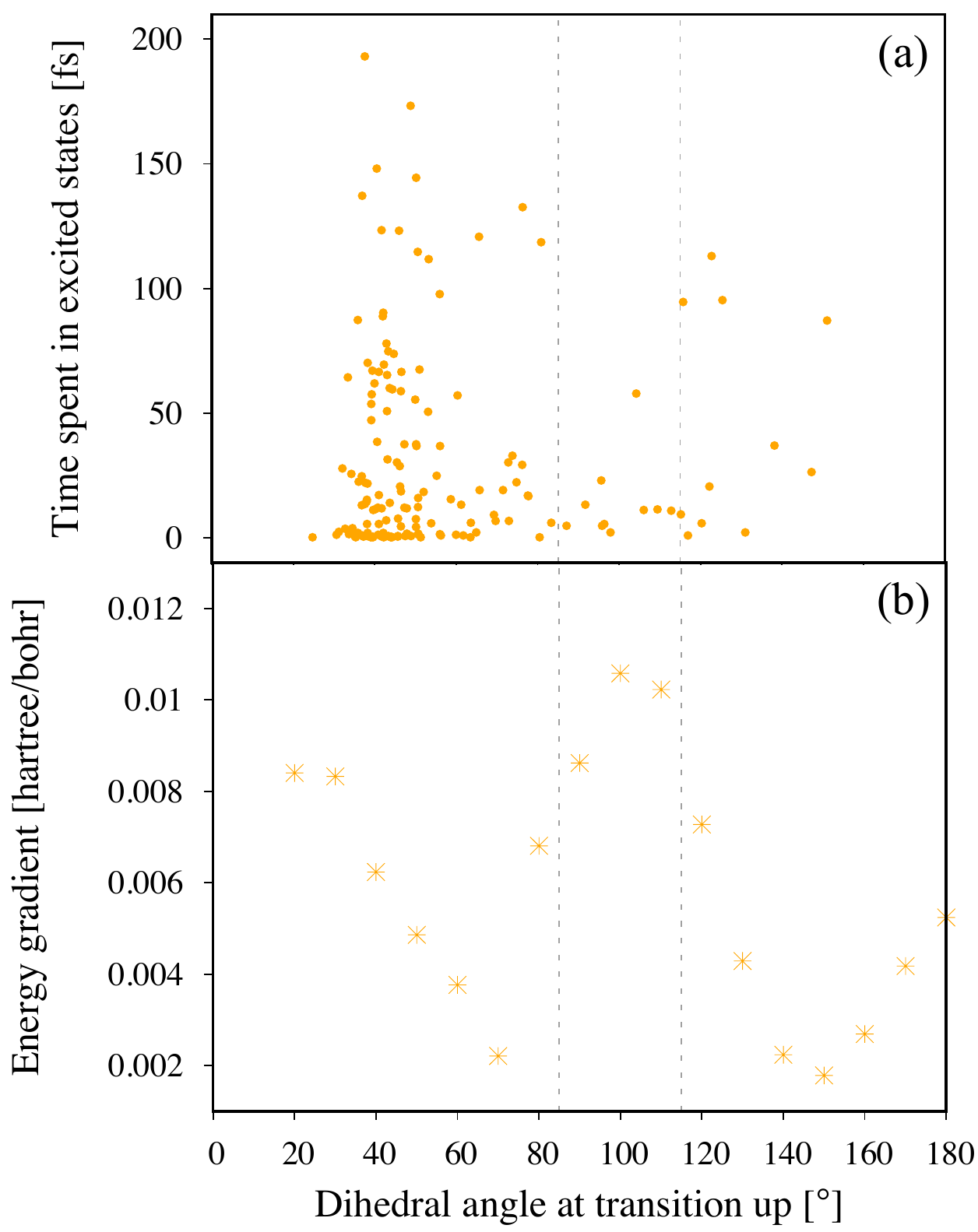}
  \caption{Dependence of the time spent in the singlet excited states and of the topology of $S_0$/$S_1$ conical intersection on the dihedral angle. (a) Time spent in singlet excited states versus dihedral angle at which the hop up occurs for the ensemble of 150 surface hopping trajectories. (b) Lowest energy gradient on the $S_1$ surface calculated at the conical intersections optimised at a range of dihedral angles.}
  \label{Fig_ExcGraDih}
\end{figure}

Over the ensemble of trajectories, the hops up to singlet excited states happen at a very wide range of geometries, illustrating that the trajectories evolve along a seam of conical intersections and can hop up at any time. Particular geometrical characteristics at which hops are more probable could not be identified.~\cite{Malhado-2016} However, there seems to be a relationship between the O-C-C-O dihedral angle at which the molecule hops up to an excited state and the time spent in the excited states upon this hop (Figure~\ref{Fig_ExcGraDih}a): long times spent in the excited states (100~fs and longer) occur when the hop up happens at a geometry with a dihedral angle between 30$^{\circ}$ and 80$^{\circ}$ and with a dihedral angle around 120$^{\circ}$ or equivalently, trajectories that have hopped up to an excited state at a dihedral angle between 90$^{\circ}$ and 110$^{\circ}$ spend a short time on the excited state before decaying back to the ground state.

In order to rationalise the dependence of the outcome of an excited state trajectory on the O-C-C-O dihedral angle of the geometry at which the hop up occurred, conical intersections between $S_0$ and $S_1$ states have been optimised for different dihedral angles (values ranging from 20$^{\circ}$ to 180$^{\circ}$) using the recent implementation of analytical non-adiabatic couplings in the Molcas package.~\cite{Galvan-2016} The optimised conical intersections have been found in the region traversed by the trajectories; they are thus relevant to the dynamics. They have been characterised using a first order approximation of the surfaces based on the energy gradients of the two states and the non-adiabatic coupling between them. For more information about the characterisation procedure used, the reader is referred to our previous work.~\cite{Galvan-2016} All optimised conical intersections are peaked, i.e.\@ they are local minima on the upper surface. They therefore act as attractors for the trajectories on the excited state. This can partly explain the fact that all trajectories eventually decay back to the ground state and none have been observed to dissociate on the excited state. At each optimised conical intersection, the smallest slope on the excited state surface which represents the ``least unfavourable way'' on the upper surface has been calculated. Figure~\ref{Fig_ExcGraDih}b plots the lowest energy gradient for the range of dihedral angles. They are always positive since the conical intersections are peaked and they vary across the range of dihedral angle illustrating the change in shape of the conical intersection along the seam. It can be noted that a path on the upper surface is more favourable at dihedral angles smaller than 90$^{\circ}$ and larger than 110$^{\circ}$. Between 90$^{\circ}$ and 110$^{\circ}$, the slope on the upper surface is larger. These results are qualitatively consistent with the dependence of the times spent in the excited states on the dihedral angle: as the slope on the excited state is greater, the trajectory decays back faster to the ground state and spends less time on the upper surface. A fair quantitative comparison is not really possible for several reasons: (i) the linear model is only valid in the close vicinity of the conical intersection; (ii) as the ensemble of trajectories evolve towards larger dihedral angles, some of them have already dissociated on the ground state, others have already hopped up to an excited state and so the number of potential trajectories that can hop up gets lower; (iii) the direction of approach of the conical intersection and the direction of lowest energy gradient also play a role in the time spent on the excited state.

In this section, non-adiabatic dynamics has been simulated, allowing non-adiabatic transitions among the four lowest-energy singlet states. It is observed that the dissociation half-life is extended to t$_{1/2}^{S}$=77~fs, compared to t$_{1/2}^{BO}$=58~fs with the ground state only. The slight ``slowing down'' of the dissociation due to non-adiabaticity effects can be further understood by a mechanism nicely presented in the experimental work of Butler~\cite{Butler-1998}. Let us explain this mechanism by considering for example the un-sampled ground state trajectory: this trajectory seems to pass in the vicinity of an avoided crossing between ground and first excited states around t=100~fs before dissociating (Figure~\ref{Fig_Ene}). When allowing for non-adiabatic effects, there is a non-negligible probability that the system behaves ``diabatically'', i.e. a non-adiabatic transition up to the excited state occurs. The trajectory would then spend some time on the excited state surface before eventual dissociation. It is unsure how this mechanism affects the overall rate of the present reaction, since the rate-limiting step is the O--O bond breaking. However, it might affect the product distribution and preferred pathways.

Also, it it noted that intersystem crossings between the singlet and triplet states have been neglected. Future dynamics simulations taking account the population of the four triplet states via spin-orbit coupling would provide complementary insights in the chemiluminescence of 1,2-dioxetane. Although the phosphorescence yield is measured to be higher than the fluorescence one, it is still very low in 1,2-dioxetane (0.3\%)~\cite{Adam-1985}. The triplet states may behave similarly to the singlet ones, i.e. they may only ``pump'' the ground state population and release it later, with very little probability of actually dissociating on the triplet state.

\section{Conclusion}
The dynamics of the decomposition reaction of 1,2-dioxetane has been simulated starting from the first O--O bond breaking transition structure. The initial geometries and velocities have been sampled from the Wigner distribution to represent the spreading of the nuclear wave packet.

The ground state decomposition of 1,2-dioxetane is a relatively fast process and the nuclear dynamics occurring during the reaction is complex. The so-called entropic trap leads to frustrated dissociations, postponing the decomposition reaction: by doing so, it determines the efficiency of the chemiluminescence. Specific geometrical conditions are necessary for the trajectories to escape from the entropic trap and for dissociation to be possible: O-C-C-O dihedral angle larger than 55$^\circ$ and O-C-C angles smaller than 117$^\circ$. Without these conditions met, the molecule stays trapped. The ground state dissociation starting from the O--O bond breaking transition structure occurs between t=30~fs and t=140~fs. The lifetime of the biradical region can be characterised by the dissociation half-time of t$_{1/2}^{BO}$=58~fs.

The singlet excited states participate as well in the trapping of the molecule by ``pumping'' the ground state population and releasing it on a 58~fs time scale approximately.  The longer the trajectory stays in the excited states, the later it dissociates. In other words, there is the traditional ``entropic trap'' mechanism occurring in the electronic ground state and in addition to it, the singlet excited states postpone further the dissociation by ``pumping up'' some population from the ground state and releasing it later. No dissociation on the singlet excited state was obtained, as expected from the extremely low singlet excitation and fluorescence yields in 1,2-dioxetane decomposition.~\cite{Adam-1985} The dissociation (including the non-adiabatic transitions to singlet excited states) now occurs from t=30~fs to t=250~fs and later. The lifetime of the biradical region is increased to t$_{1/2}^{S}$=77~fs.

The hops of the trajectories up to an excited state occur at a wide range of the geometries along the seam of conical intersections. There is a connection between the time spent by a trajectory on the excited states after a hop and the dihedral angle of the geometry at which the hop happens. To explain this, several conical intersections between $S_0$ and $S_1$ states along the seam (with various dihedral angles) have been optimised and characterised. Although all conical intersections are peaked, their topography varies with the dihedral angle. For instance, the conical intersection optimised at a dihedral angle of 50$^{\circ}$ presents a small slope on the upper surface and thus allows the trajectory to ``wander'' for longer before going back to the ground state, while the conical intersection optimised at a dihedral angle of 100$^{\circ}$ presents a greater slope on the upper surface and thus attracts back the trajectory faster.

Future dynamics simulations on substituted dioxetanes with various chemiluminescence quantum yields would help investigating further how the entropic trap determines the efficiency of the chemiluminescent process. Inclusion of the triplet states in the simulation would also allow more direct comparison between the experimental excitation yields and the simulated final populations of singlet and triplet excited states.

\begin{acknowledgement}

This work was supported by  the Swedish Research Council (Grant No.\@ 2012-3910). The simulations were performed on resources provided by the Swedish National Infrastructure for Computing (SNIC) at UPPMAX and NSC centers.

\end{acknowledgement}

\begin{suppinfo}
Comparison between CASSCF and CASPT2 potential energy cuts and single point MS-CASPT2 potential energy evaluations of the four lowest-energy singlet states along the un-sampled trajectory. Study of the dependence of the results with respect to the dissociation threshold. Transition state structure and initial nuclear velocities along the transition vector.
\end{suppinfo}

\bibliography{references}

\begin{mcitethebibliography}{45}
\bibitem[Not()]{Note-2}
ANO basis sets are systematically improvable, and produce smaller basis set
  superposition errors. Also, the ANO-RCC basis set was developed using
  correlated methods and aimed at providing a balanced description of various
  electronic states.\relax
\mciteBstWouldAddEndPunctfalse
\mciteSetBstMidEndSepPunct{\mcitedefaultmidpunct}
{}{\mcitedefaultseppunct}\relax
\EndOfBibitem
\bibitem[Not()]{Note-3}
Frustrated dissociations lead to elongation of the C--C bond up to 1.8~\AA. To
  make the comparison between the geometries of frustrated versus successful
  dissociations fair, dissociation is here considered to happen when the C--C
  bond length exceeds 2.0~\AA~(instead of 2.4~\AA).\relax
\mciteBstWouldAddEndPunctfalse
\mciteSetBstMidEndSepPunct{\mcitedefaultmidpunct}
{}{\mcitedefaultseppunct}\relax
\EndOfBibitem
\bibitem[Not()]{Note-5}
The fact that the rise time $\tau$ is equal to the ground state dissociation
  half-life t$_{1/2}^{BO}$ is obviously a coincidence.\relax
\mciteBstWouldAddEndPunctfalse
\mciteSetBstMidEndSepPunct{\mcitedefaultmidpunct}
{}{\mcitedefaultseppunct}\relax
\EndOfBibitem
\end{mcitethebibliography}

\end{document}